\documentclass[aps,twocolumn,superscriptaddress,preprintnumbers,floatfix,nofootinbib]{revtex4}
\usepackage{epsfig,comment}
\usepackage{graphicx,amssymb,epsfig,amsmath}
\usepackage{color}

\newcommand{\be}{\begin{eqnarray}}
\newcommand{\ee}{\end{eqnarray}}

\newenvironment{proof}[1][Proof]{\textbf{#1.} }{\ \rule{0.5em}{0.5em}}

\newtheorem{lem}{Lemma}
\newtheorem{trm}{Theorem}

\begin{document}

\title{Robust amplification of Santha-Vazirani sources with three devices}

\author{Piotr Mironowicz}
	\affiliation{Department of Algorithms and System Modelling, Faculty of Electronics, Telecommunications and Informatics, Gda\'{n}sk University of Technology, Gda\'{n}sk 80-233, Poland}
	\affiliation{National Quantum Information Centre in Gda\'{n}sk, Sopot 81-824, Poland}

\author{Rodrigo Gallego}
	\affiliation{Dahlem Center for Complex Quantum Systems, Freie Universitaet Berlin, 14195 Berlin, Germany}

\author{Marcin Paw\l{}owski}
	\affiliation{Department of Mathematics, University of Bristol, Bristol BS8 1TW, U.K.}
	\affiliation{Institute of Theoretical Physics and Astrophysics, University of Gda\'nsk, 80-952 Gda\'nsk, Poland}

\date{\today{}}

\begin{abstract}
We demonstrate that amplification of arbitrarily weak randomness is possible using quantum resources. We present a randomness amplification protocol that involves Bell experiments. We find a Bell inequality which can amplify arbitrarily weak randomness and give a detailed analysis of the protocol involving it. Our analysis includes finding a sufficient violation of Bell inequality as a function of the initial quality of randomness. It has a very important property that for any quality the required violation is strictly lower than possible to obtain using quantum resources. Among other things, it means that the protocol takes a finite amount of time to amplify arbitrarily weak randomness.
\end{abstract}

\maketitle

\section{Introduction}

The application of the laws of quantum mechanics allows to perform tasks impossible in classical information theory. The two most prominent examples are quantum computation \cite{QComp} and cryptography \cite{review}. Recently, another area where quantum information theory makes new things possible has been found. It is the amplification of weak randomness \cite{Amp1}. This procedure not only has obvious practical applications, but it also sheds a light on fundamental issues such as completeness of quantum mechanics. However, so far the possibility of randomness amplification has been demonstrated only under very restrictive conditions. To explain what they are we must first rigorously state the problem.

We are given a source, which generates a sequence of bits $\vec{x}=x_0,x_1,...$ parameterized by a single constant $\epsilon$. The bits may be correlated with each other and also with an agent, the eavesdropper, that holds a classical variable $e$. However, there is a certain intrinsic randomness in each of the bits quantified by $\epsilon$ in
\be
	\label{svcon}
	\forall_i \quad \frac{1}{2}-\epsilon \leq P(x_i=0|x_0,...,x_{i-1},e) \leq \frac{1}{2}+\epsilon.
\ee
If (\ref{svcon}) holds we say that the sequence $\vec{x}$ (or the source) is $\epsilon$-free. $\epsilon$ is called the bias of the source. $\epsilon=0$ corresponds to the case where the output of the source is perfectly random. When $\epsilon=\frac{1}{2}$ we cannot say anything about the source and it can be even deterministic. The aim of randomness amplification is to use some postprocessing of the sequence $\vec{x}$ to generate another sequence $\vec{y}$ which is $\epsilon'$-free and $\epsilon'<\epsilon$.

The source of randomness described above is usually referred to as a Santha-Vazirani source after the authors of \cite{SV}, where they have proved that classical randomness amplification is impossible. In their groundbreaking paper Colbeck and Renner \cite{Amp1} showed that it is not true in the quantum case. Their idea is based on performing Bell experiment and applying a hashing function to the measurement outcomes. However, the protocol that they have presented works only if the source of randomness is almost perfect to begin with. More precisely: if the source is $\epsilon$-free with $\epsilon<0.086$.

Recently, more papers on this issue appeared, aiming at amplification of any source with $\epsilon<\frac{1}{2}$. Unfortunately, they either work only in the noiseless case\cite{Amp2}, which is impossible in realistic experimental situations; require unbounded number of devices \cite{Amp3} or have a zero rate of amplification \cite{Amp4,Amp5}. These protocols assume only no--signaling, but for the reasons mentioned fail to perform practically usable amplification.

Therefore, it remained an open question whether the amplification of arbitrarily weak randomness under realistic circumstances was possible. In this paper we answer this question affirmatively. We do so by presenting an amplification protocol which is based on Mermin inequality \cite{Mermin}. It works for any $\epsilon<\frac{1}{2}$ and can tolerate a finite amount of noise and experimental imperfections depending on $\epsilon$.

The aim of the paper is not to study the foundations of the quantum theory but the details of amplifying randomness in practice. Therefore, contrary to the majority of papers on the subject \cite{Amp1,Amp2,Amp3,Amp4,Amp5,Amp6}, we require the vendor of the devices to be bound by laws of quantum mechanics instead of only no-signalling. This enables us to develop a simple, noise tolerant protocol with only a few, three to be precise, reusable devices\footnote{By reusable we mean devices which can be used many times in a single run of the protocol.}.

Recently, a protocol which is able to amplify any randomness (not only from Santha-Vazirani sources) was proposed \cite{CSW}. While it it clearly more general than ours it is not specified how much experimental imperfections can be tolerated and the number of devices required for the operation is of the order of $10^7$ \cite{A-son} which is much more than 3 devices required by ours.

\section{The task of randomness amplification}

The main problem with randomness amplification lies in our almost complete ignorance about the inner workings of the source. It provides us with an infinite sequence of bits, yet all we know about it is expressed by a single number $\epsilon$. For every $\epsilon$ there exists an infinite number of $\epsilon$-free sources and, while good randomness amplification procedures would work well for a vast majority of them, there will always be some that any given procedure fails to amplify. This is the essence of Santha-Vazirani proof.

Another problem is that, since we are interested only in the quality of the sequences, we do not have access to any independent source of randomness. Or, in other words, we assume that all the sources of randomness that we have access to can be correlated and form one big Santha-Vazirani source. Therefore, without loss of generality, any classical randomness amplification protocol can be reduced to applying a deterministic function to the output of the source to generate a new sequence: $\vec{y}=f(\vec{x})$.

Can quantum mechanics help? After all it is a theory built on intrinsic randomness. On the other hand, we cannot simply use a quantum random number generator because, under our assumptions, it is also only $\epsilon$-free. The solution lies in Bell inequalities. They have already been found useful in a related problem of device independent randomness expansion \cite{rndexp1,rndexp2,rndexp3,rndexp4,rndexp5,rndexp6}. In a nutshell, the idea is to use the sequence $\vec{x}$ to choose the settings of a Bell experiment and consider the outcomes as your new sequence $\vec{y}$. Based on these sequences the violation of the Bell inequality is estimated, and its value tells us if the amplification was successful or not. Let us elucidate.

\subsection{Biased nonlocal games}
One way of interpreting Bell inequalities is to think of them as nonlocal games. Let us take CHSH \cite{CHSH} as an example. If we treat it as a game then we have a team of two players, Alice and Bob, playing against a referee. The referee sends bits $a$ and $b$ to Alice and Bob, respectively, and the players, without communication, announce their respective binary answers $A$ and $B$. They win if $A\oplus B=ab$. Usually, it is assumed that the probability distribution of the inputs is uniform, \textit{i.e.} $p(a,b)=\frac{1}{4}$. Under this condition the maximal winning probability for the parties having only classical resources is $\frac{3}{4}$, while entanglement allows them to reach the success probability up to $\frac{1}{2}\left(1+\frac{1}{\sqrt{2}} \right)$.

If we are to use $\epsilon$-free string of bits as a source of settings, we cannot assume that the distribution of inputs is uniform anymore. Our game becomes a biased one, at least from the eavesdroppers point of view, with the success probability
\be
	\label{eq:chsh}
	P_s=\sum_{a,b}p(a,b|e)P(A\oplus B=ab|a,b,e),
\ee
where $e$ is a variable held by the eavesdropper. The values of $P_s$ for classical and quantum strategies for any distribution $p(a,b|e)$ are greater than their counterparts from the unbiased case. They have been found in \cite{biased}. For every Bell inequality, the larger the observed value of $P_s$ is the more random the local outcomes must be. One can use the hierarchy of semi-definite programs (SDPs) from \cite{NPA} to efficiently find a lower bound on this randomness for any given distribution $p(a,b|e)$. Unfortunately, we do not know this probability distribution. We cannot even estimate it because it may be different in each round of the experiment and the choice of the distribution may be correlated with variable $e$ held by the eavesdropper. The only thing that we know about $p(a,b|e)$ is that both bits $a$ and $b$ come from an $\epsilon$-free source. But the impossibility of the direct application of SDP is not the last of the obstacles.

Let us assume that our $\epsilon$-free source is always biased towards 0, \textit{i.e.} for all $i$ $P(x_i=0|x_0,...,x_{i-1},e) = \frac{1}{2}+\epsilon$. Even if Alice and Bob know this and adopt their states and measurements accordingly, there is a value of $\epsilon$ above which quantum and classical $P_s$ are the same. One can use the results from \cite{biased} to find this value to be $\epsilon_{crit}=\frac{1}{\sqrt{2}}-\frac{1}{2}$. And if a classical model that gives certain success probability in a nonlocal game exists, then there is also a deterministic one achieving it \cite{Fine}. Therefore, whenever the parties have a source with $\epsilon$ above $\epsilon_{crit}$, whatever success probability they observe their outcomes can be deterministic. Fortunately, the value of $\epsilon_{crit}$ depends on the Bell inequality chosen for the protocol. Therefore, our task is also to find Bell inequalities which are better for the purposes of randomness amplification than CHSH or the ones studied in \cite{Amp1}.

\subsection{Protocol: notation and assumptions}

The protocol of amplification of weak randomness considers two devices: the source of randomness (SoR) and the quantum box (QB). These two devices are operated by honest players, however they may have been manufactured by dishonest agents as long as they fulfill the assumptions that we make explicit further. The task of randomness amplification is performed in competition with a dishonest player, the eavesdropper, having a device that generates the random variable $e$. The goal is to generate a final bit $y$ that is $\epsilon'$-free with respect to the eavesdropper, that is $\frac{1}{2}-\epsilon' \leq P(y|e) \leq \frac{1}{2} + \epsilon'$.

In the scenario that we are considering, the eavesdropper, after preparing SoR and QB is quite passive. Whole randomness amplification procedure is performed in a shielded lab (see Fig. 3) and no information is sent outside. The eavesdropper does not even know at which time the protocol is run. This means that if she has any quantum side information (a system entangled with the devices she has produced) she learns the same amount if she measures her system after or before the protocol. In fact nothing changes even if she measures it before she gives the devices to the players. Therefore, in this scenario there is no difference between eavesdroppers with classical and quantum side information and our protocol is secure against both.

Note, that we differ to cryptographic scenario in the point that, after the amplification process, we do not need to keep the generated bits secret, since we are interested only in their indeterminacy, not privacy.

The SoR is assumed to fulfill the assumptions of a Santha-Vazirani source, that is, the honest players can produce an arbitrarily large number of bits $\vec{x}$ according to a probability distribution that fulfills \eqref{svcon}. The QB receives as inputs $k$ classical bits and produces $k$ classical bits. The QB is reused an arbitrarily large number of times. Let us denote by $\vec{a}_j\in \{0,1\}^k$ and $\vec{A}_j\in\{0,1\}^k$ the inputs and outputs of the $j$-th run, respectively\footnote{In the particular scenario of $k=2$ or $k=3$ we employ for ease of notation $a_j,b_j,c_j$ and $A_j,B_j,C_j$ as inputs and outputs of the $j$-th run, respectively.}. Let us also denote by $\vec{X}_j$ all the inputs and outputs generated until the $j$-th run. The behavior of the QB in the $j$-th run on a given instance of $e$, $\vec{x}$ and $\vec{X}_{j-1}$ is determined by the probability distribution $P(\vec{A}_j|\vec{a}_j,\vec{x},\vec{X}_{j-1},e)$. We denote the success probability of the $j$-th round as
\be
	\label{eq:psj}
	P_s^j = \sum_{\vec{a}_j} P\left(F[\vec{a}_j,\vec{A}_j]=0 \bigg| \vec{a}_j,\vec{X}_{j-1},\vec{x},e\right) P(\vec{a}_j|\vec{X}_{j-1},e,\vec{x}),
\ee
where $F[\vec{a}_j,\vec{A}_j]=0$ is the function that determines the Bell inequality employed, \textit{cf.} \eqref{eq:chsh}. Also, we refer to the average success probability of $N$ runs of the QB to $P_{\text{ave}}:=\frac{1}{N}\sum_{j=1}^N P_s^j$.

Now we enunciate the two assumptions on the QB:
\begin{itemize}
\item[i)] \emph{Markov condition:} We assume that given a value of $e$ and the inputs $\vec{a}$, the outputs probability distribution of the QB is independent of the bits generated by SoR. That is
	\begin{eqnarray}
		\nonumber &&P(\vec{A}_1,...,\vec{A}_N|\vec{a}_1,...,\vec{a}_N,\vec{x},e)\\
		\label{eq:markov}&=&P(\vec{A}_1,...,\vec{A}_N|\vec{a}_1,...,\vec{a}_N,e).
	\end{eqnarray}
	Note that this assumption does not imply at all that SoR and QB are uncorrelated. It just states that they have to be correlated only through the random variable $e$ possessed by the eavesdropper. In this way, the QB is for every $e$ a well-defined channel applicable to the bits $\vec{x}$ generated by SoR. Indeed, in our protocol the inputs $\vec{a}$ are generated by the SoR, so that $\vec{a}=(x_1,...,x_N)$. Let us denote by $\vec{h}$ the rest of the bits generated by the SoR that are not explicitly used as inputs of QB. Then the Markov condition \eqref{eq:markov} implies that
	\begin{eqnarray}
		\nonumber &&P(\vec{A}_1,...,\vec{A}_N,\vec{h}|\vec{a}_1,...,\vec{a}_N,e)\\
		\label{eq:markov2}&=&P(\vec{A}_1,...,\vec{A}_N|\vec{a}_1,...,\vec{a}_N,e)\times P(\vec{h}|\vec{a}_1,...,\vec{a}_N,e).
	\end{eqnarray}
\item[ii)] \emph{Quantum behavior:} As stated in the introduction, we assume that the devices fulfill the rules of quantum mechanics. Therefore we assume that for all $j$ there exist a $k$-partite quantum state $\rho(e\vec{X}_{j-1})$ and measurement operators $M_{\vec{A}_j}^{\vec{a}_j}(e,\vec{X}_{j-1})\equiv \bigotimes _{i=1}^{k}M_{A_j^i}^{a_j^i}(e,\vec{X}_{j-1})$, with $\sum_{A_j^i} M_{A_j^i}^{a_j^i}(e,\vec{X}_{j-1})=\mathbb{I}$ for all $i$, such that
	\begin{equation}
		P(\vec{A}_j|\vec{a}_j,\vec{X},e)= \text{tr}\left(\rho(e\vec{X}) M_{\vec{A}_j}^{\vec{a}_j}(e,\vec{X})\right)
	\end{equation}
\end{itemize}

\subsection{Sketch of the protocol and proof}

The protocol of randomness amplification is based on the fact that a probability distribution with a sufficiently large success probability for a certain Bell game can be certified to posses some intrinsic randomness. We employ the SoR to generate the inputs of the QB at every run. The remaining bits generated by SoR are referred to as $\vec{h}$. The behavior of the $j$-th run of the QB box is characterized by $P(\vec{A}_j|\vec{a}_j,\vec{X}_{j-1},e)$ (note that it does not depend on $\vec{h}$ due to the \emph{Markov assumption}). In Sec. \ref{sec:bellinequalities} we show that there exists a Bell inequality such that one of the bits produced by the QB can be shown to be $\epsilon_j$-free, with $\epsilon_j$ being a function of $P_s^j$ and $\epsilon$. More precisely
\be
\frac{1}{2}-\epsilon_j \leq P(A_j^1|\vec{a}_j,\vec{X}_{j-1},e)\leq \frac{1}{2}+\epsilon_j
\ee
with $\epsilon_j=g(P_s^j,\epsilon)$, where we have chosen the first of $k$ bits, $A_j^1$, generated by the QB in $j$-th run.

Unfortunately, the success probability $P_s^j$ of each run cannot be estimated. By using the QB $N$ times, we have only access to one event of each run; hence we deal with the estimated success probability $P_{\text{est}}$ (the number of runs that won the game divided by the number of runs, $N$). We cannot employ standard estimation results in nonlocality because now the distribution of measurement settings is unknown. The standard scenario\cite{rndexp2,rndexp5,rndexp6} assumes that it is possible to use an estimator of the form
\be
	\hat{I} = \frac{1}{N} \sum_{j=1}^{N} \sum_{\vec{A},\vec{a}} c_{\vec{A},\vec{a}} \frac{\chi(\vec{A},\vec{a})}{P(\vec{a})}.
\ee
This time the values of $P(\vec{a})$ can differ in each round in a way that cannot be predicted without the knowledge of internal working of SoR.

But we can bound the winning probability of a virtual game: an unbiased one played with the same states and measurements and from it obtain the bounds on the average bias of the $N$ bits, $\epsilon_{\text{ave}}:=\frac{1}{N}\sum_{j=1}^N \epsilon_j$.

Clearly, bounds on the average bias of all the bits generated does not complete the proof. One may have very good bounds on the average bias, however some fraction of the bits $\{A_j^1\}_j$ may not be random at all. Hence, one cannot certify that one of them --chosen at random employing the SoR-- will posses any randomness. This approach (of using SoR to chose some of the output bits) was used in \cite{Amp1,Amp2,Amp6}. It was slightly modified in \cite{Amp3} where a constant number of rounds is chosen and XOR of the outcomes is the final value. However, if the devices are allowed to have even tiniest imperfections, this strategy (i.e. using SoR to pick a constant number of bits) cannot amplify sources with $\epsilon \geq \frac{2e-1}{4e+2} \approx 0.345$ regardless of Bell inequality chosen and the level of imperfections. This is shown in the appendix by providing an explicit classical attack. In this work, we employ a more sophisticated post processing based on techniques in Ref. \cite{chor} that allows one to extract a fully random bit.

\section{Bell inequalities}
\label{sec:bellinequalities}

We are considering a protocols of a certain structure, where the estimate of a winning probability in some non-local game is the only parameter used to check f the amplification was successful. To find a candidate for a Bell inequality to be used in randomness amplification protocol we first need to ask ourselves what properties are we looking for. To this end, let us consider one particular way of cheating\footnote{By cheating we here mean violating Bell inequality with deterministic outcomes.}. The measurement devices prepare an optimal classical strategy so they know in advance that for most of the messages from the referee they will produce a good answer, but they also know that for some they will fail. The devices also know for which inputs they will fail. They can be tuned to the source of randomness in such a way that the inputs for the case when the devices fail are least likely to happen. We see that the weaker the randomness the higher the average success probability. If the randomness is very weak, the success probability is close to 1. If we want to amplify arbitrarily weak randomness, then to be certain that our device does not play this trick, the success probability with a quantum strategy has to be even higher. Obviously, it cannot be greater than 1, so we have to look for nonlocal games for which it is equal to 1.

Taking this all into account we choose the tripartite Mermin inequality \cite{Mermin} as our candidate. In this scenario Alice, Bob and Charlie each receive one input bit, $a,b$ and $c$, respectively. There is a promise that $a\oplus b\oplus c=1$. Each of them also returns a single bit denoted $A,B$ and $C$. They win if $A\oplus B\oplus C=abc$. In the unbiased version of this game the classical success probability is $\frac{3}{4}$, and quantum mechanics allows to reach 1.

\section{Bounding the randomness}

As we mentioned before, finding lower bounds on the quality of randomness generated by playing a biased nonlocal game is highly non-trivial.

Because of the promise put on the choices of the settings one can write the success probability of a nonlocal biased game based on the tripartite Mermin inequality as
\be
	P_s=\sum_{a,b} p(a,b|e)P(A\oplus B\oplus C=ab|a,b,e)
\ee
Our aim is to find an upper bound on the following quantity
\be
	P_{max}=\max_{a,b,c,X,i,e}P(X=i|a,b,c,e),
\ee
as a function of $\epsilon$ and $P_s$ under constraints
\be
	\label{nonlin}
	\sum_{a,b} p(a,b|e)P(A\oplus B\oplus C=ab|a,b,e)\geq P_s
	\\
	\frac{1}{2}-\epsilon\leq p(a|e)\leq\frac{1}{2}+\epsilon
	\\ \label{nonlin2}
	\frac{1}{2}-\epsilon\leq p(b|a,e)\leq\frac{1}{2}+\epsilon
\ee
where $X\in\{A,B,C\}$ denotes the outcome of one of the parties. Because the conditions (\ref{nonlin}-\ref{nonlin2}) are nonlinear we cannot use semi-definite programming to solve this problem. Moreover, we are not able to construct the expression $\sum_{a,b} p(a,b|e)P(A\oplus B\oplus C=ab|a,b,e) \geq P_s$, since the values $p(a,b|e)$  are not constant (in contrast to standard scenario with SoR without memory).

To cope with it, let us consider what would the success probability be if an unbiased game was played using the same states and measurements. Let us denote it by
\be
P_s^{ub}=\frac{1}{4}\sum_{a,b}P(A\oplus B\oplus C=ab|a,b,e).
\ee
The fact that the same states and measurements are used for this virtual game means that the probabilities $P(A\oplus B\oplus C=ab|a,b,e)$ in the formula above are exactly the same as in (\ref{nonlin}). We can use this to bound $P_s^{ub}$ from below. Let $P_m=\min_{a,b} P(A\oplus B\oplus C=ab|a,b,e)$. We have that
\be
P_s^{ub}\geq P_m.
\ee
The highest value of $P_s$ which is still consistent with $P_m$ is attained if $P(A\oplus B\oplus C=ab|a,b,e)=1$ for all $a$ and $b$ except for the ones used to get $P_m$. Moreover the coefficient $(a,b|e)$ in front of $P_m$ should be as small as possible for something coming out of a Santha-Vazirani source parameterized by $\epsilon$, i.e. $\left(\frac{1}{2}-\epsilon\right)^2$. This implies
\be \nonumber
P_s\leq 1-\left(\frac{1}{2}-\epsilon\right)^2+\left(\frac{1}{2}-\epsilon\right)^2 P_m
\\
\leq 1- \left(\frac{1}{2}-\epsilon\right)^2\left(1-P_s^{ub}\right)
\ee
or, alternatively,
\be \label{est}
P_s^{ub}\geq 1-\frac{1-P_s}{\left(\frac{1}{2}-\epsilon\right)^2}
\ee

Now we can bound $P_{max}$ by considering only the constraint
\be
\frac{1}{4}\sum_{a,b}P(A\oplus B\oplus C=ab|a,b,e)\geq P_s^{ub}
\ee
and using (\ref{est}) to lowerbound $P_s^{ub}$ by $P_s$ which is a quantity that we can experimentally estimate.

We find that one can obtain good bounds on $P_{max}$ already with the first intermediate level of the hierarchy $Q_{1+AB+AC+BC}$ from \cite{NPA}. The main result of our analysis so far is that for any\footnote{In fact we have numerically checked only $\epsilon \leq 0.499$, but conjecture this is true for all $\epsilon < \frac{1}{2}$.}  $\epsilon < \frac{1}{2}$ there exists $P_s<1$ such that $P_{max}<1$. A function $g(P_s,\epsilon)$, giving a concave upper bound on $\tilde{P}_{max}$, and thus on $P_{max}$, is plotted in Fig.1. The critical value of $P_{crit}(\epsilon)$ such that for all $P_s>P_{crit}(\epsilon)$ we have $g(P_s,\epsilon)<\frac{1}{2}$ is shown in Fig.2.

\begin{figure}[!htbp]
	\center
	\resizebox{8cm}{!}{
	\includegraphics{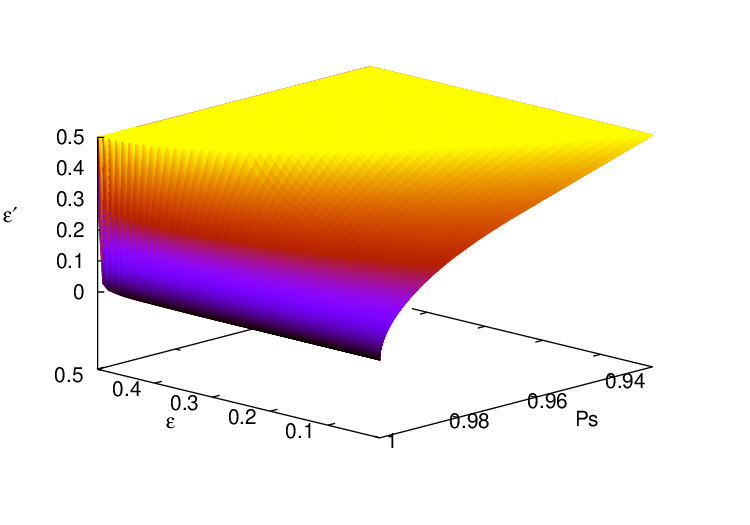}}
	\caption{(Color online) Maximal bias of the measurement outcome as a function of the weakness of randomness and success probability of winning in a nonlocal game based on tripartite Mermin inequality. This plot can be also understood as a critical value of the success probability required to take inputs from $\epsilon$-free source and get outcomes with bias less than $\epsilon'$. }
\end{figure}

\begin{figure}[!htbp]
	\center
	\resizebox{8cm}{!}{
	\includegraphics{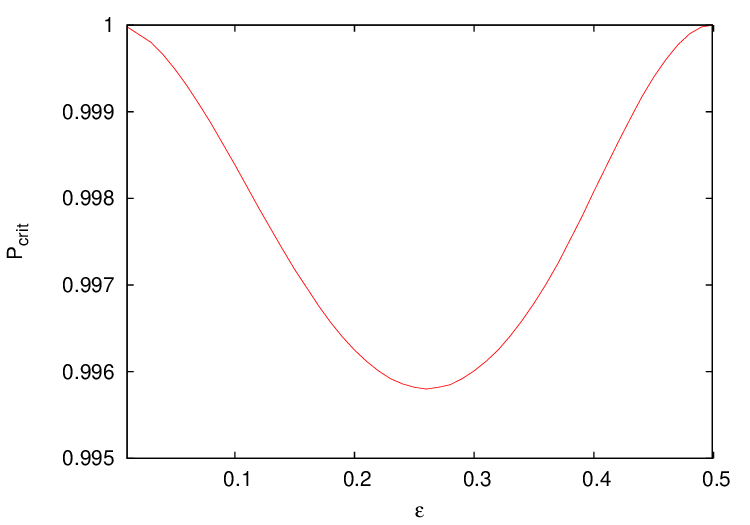}}
	\caption{(Color online) Sufficient winning probability for randomness amplification in a nonlocal game based on tri-partite Mermin inequality as a function of the sources' freedom. If $P_s$ is above the plot it means that the bias of the final bit $y$ is lower than that of the SV source. Note that higher values of $P_s$ are required for extremal initial $\epsilon$'s. This is because if $\epsilon$ is large the initial quantity of randomness is bad which makes the amplification difficult. On the other hand if $\epsilon$ is already low then to amplify it we need to obtain an even more random bit which is again difficult.}
\end{figure}

\section{Full randomness amplification protocol}

If, in every round of the experiment, the device would have the same probability of success, we could simply take the outcomes of the experiment as our final sequence $\vec{A}$ and know that each of them is $\epsilon'$-free with $\epsilon' = g(P_s,\epsilon)$. Unfortunately, we cannot assume that, and have to apply some classical postprocessing. We also cannot randomly choose a single bit from the measurements outcomes like it was done in \cite{Amp1} because for such postprocessing amplification of randomness with $\epsilon>0.345$ is impossible regardless of the Bell inequality chosen as a certificate \footnote{This claim is proved in the appendix C.}. Therefore, we need to use a different protocol but before we present it let us clarify the task at hand.

We are given a device which is a source of randomness guaranteed to be $\epsilon$-free. We assume that the vendor of the device, that may be the eavesdropper himself, has access to some parameters $e$ which influences its behavior. Moreover, any other device that we have access to (\textit{e.g.} a source of entangled states) is also supplied by the same vendor and its behavior is also dependent on $e$. We are able to place the source of randomness and all the other devices we need in a lab shielded from the environment in such a way that it leaks no data to the outside world during the amplification process, especially to the said vendor. In this lab a single bit $y$ is generated \footnote{If we have a procedure that allows us to generate a single bit with required $\epsilon'$-freedom, then we can repeat the same procedure any number of times, including the bits already obtained in $e$ to get a sequence of any length.}. We choose some target $\epsilon'<\epsilon$ and a desired probability of success $p$ in generating a final bit that fulfills
\be
\frac{1}{2}-\epsilon'\leq P(y=0|e)\leq \frac{1}{2}+\epsilon'.
\ee

To this end, we use the following amplification protocol:
\begin{enumerate}

	\item Place in a shielded lab an $\epsilon$-free Source of Randomness (SoR) and a Quantum Box (QB). The latter is composed of measurement devices sharing entangled state, designed for demonstrating violation of the tripartite Mermin inequality.

	\item Use SoR to draw $2N$ bits for the measurement settings for QB for $N$ rounds of the experiment and make the measurements.

	\item Use SoR to draw $N$ bits representing a hashing function. Label them $h_1,...,h_N$.

	\item Calculate $y=\bigoplus_{i=1}^N h_i A_i$ where $A_i$ is the outcome of Alice in $i$-th round. Then estimate the average Bell inequality violation $P_{est}$ of QB to compute the probability that $y$ is $\epsilon'$-free. If it fulfills
		\begin{equation}
			\label{eq:estsuccess}
			g\left(P_{\text{est}}-r(p,N,\epsilon), \epsilon\right)<\frac{1}{1+2\epsilon}-\frac{1}{2}
		\end{equation}
		with $r(p,N)=\sqrt{-\ln(1-p)}N^{-\frac{1}{2}}$ and $g$ from Fig. 1; keep $y$, otherwise abort.
	
\end{enumerate}

The protocol is schematically pictured in Fig.3. The high probability mentioned in the last step is made explicit by the following theorem:
\begin{trm}
	For any $\epsilon<\frac{1}{2},\epsilon'>0$ and $p<1$ there exists $N$ such that in the protocol presented above the bit $y$ is $\epsilon'$-free with probability at least $p$ when $g(P_{est},\epsilon)>\frac{1}{1+2\epsilon}-\frac{1}{2}$.
\end{trm}
\begin{proof}
	A proof is given in the appendix.
\end{proof}

\begin{figure}[!htbp]
	\center
	\resizebox{9cm}{!}{\includegraphics{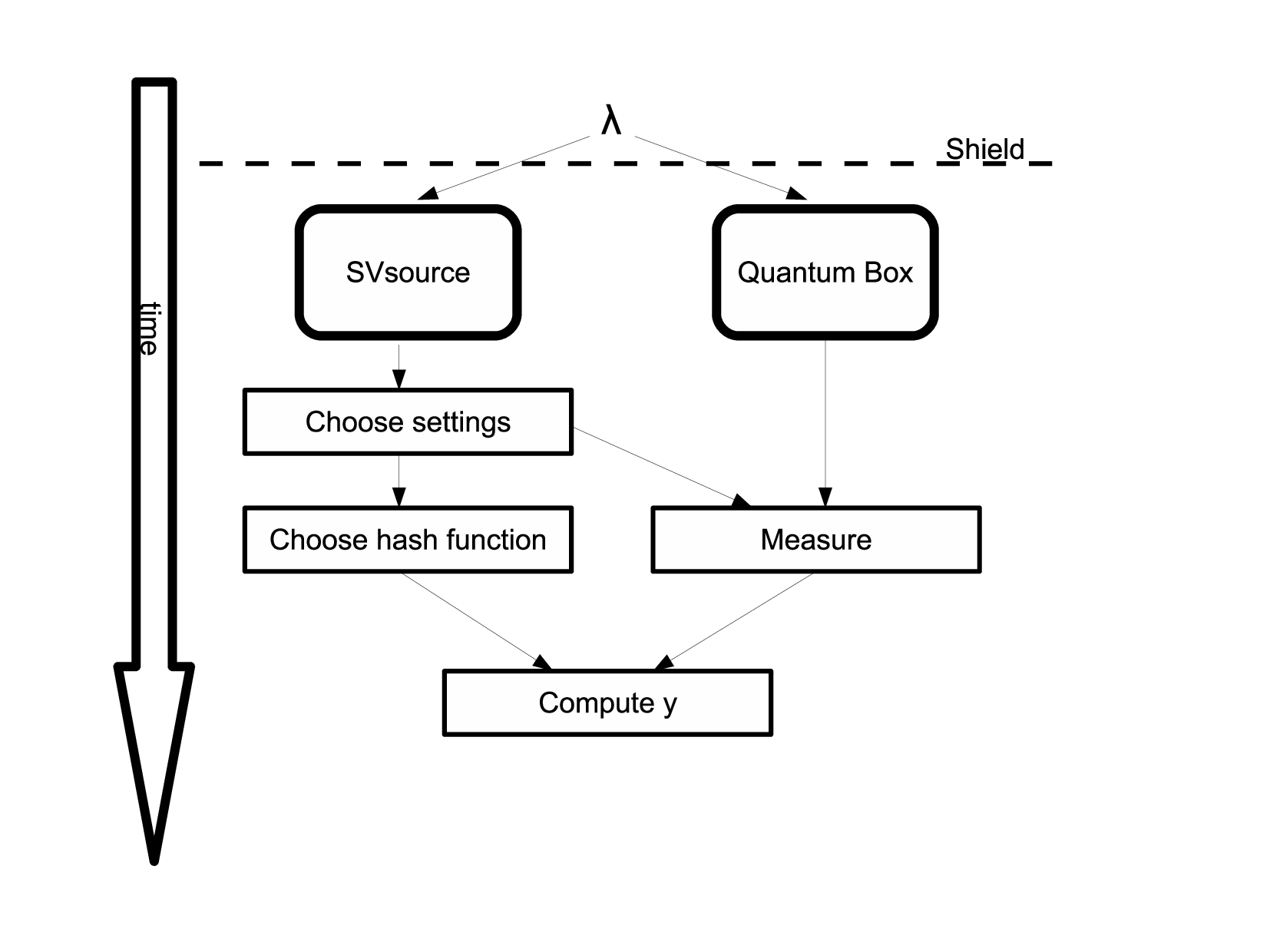}}
	\caption{Schematic diagram of the randomness amplification protocol.}
\end{figure}

\section{Conclusions}

We have demonstrated that amplification of arbitrarily weak randomness is possible using quantum resources. We were able to derive the necessary bounds on the violation of Bell inequalities as a function of the randomness' quality. These bounds are below the maximum achievable by quantum resources for arbitrarily weak initial randomness. We have also presented a protocol that uses Bell inequalities for randomness amplification and calculated all of its parameters.

We find of a particular interest the fact that the inequality we have demonstrated to perform well in randomness amplification, does so only in quantum theory, it cannot amplify any randomness of any quality when only no-signalling is assumed, because the adversary, though unable to make the whole outputs of all the parties deterministic, can always fix any single bit of it \cite{Amp2}. Although a different protocol based on the same inequality might work we find it interesting that for the one presented here the difference between quantum and no-signalling theory is qualitative rather than quantitative. This result is in contrast with protocols based on other Bell inequalities where the theory of the adversary only influences the efficiency of the protocol but not the possibility of its secure execution \cite{MAG}.

As our paper solves an open problem it also poses some new ones:  Are there any protocols more efficient than ours in the terms of ratio of random bits generated to random bits used? What is the lowest violation of Bell inequality required for amplification of a certain $\epsilon$-source with three devices?

\acknowledgments

This work is supported by UK EPSRC, FNP TEAM, NCN grant 2013/08/M/ST2/00626, ERC QOLAPS and IDEAS PLUS (IdP2011 000361). SDP was implemented in OCTAVE using SeDuMi toolbox \cite{SeDuMi}.

\begin{widetext}

\section{Appendix A: Proof of Theorem 2}

After performing steps 1.-4. one has $N$ bits $A_1,\ldots,A_N$ generated by the QB and $N$ bits $h_1,\ldots,h_n$ generated by SoR. The rest of the bits generated in the process --such as bits for the measurement settings or the QB outputs that are NOT employed to construct $A_1,...,A_N$-- are referred to as $\vec{X}$ for ease of notation. To summarize, let us recall the three elements that will come into play in the following proof:

\begin{enumerate}

	\item By the assumption on the behavior of the SoR, bits $h_1,...,h_N$ fulfill
		\begin{equation}
			\frac{1}{2}-\epsilon \leq P(h_i|h_1,...,h_{i-1})\leq \frac{1}{2}+\epsilon \:\:\: \forall i\leq N \nonumber
		\end{equation}
	
	\item The observed statistics reveal an estimated success probability $P_{\text{est}}$ --calculated simply as the number of rounds when the parties won, divided by the total number of rounds $N$-- that fulfills equation (\ref{eq:estsuccess}), \textit{i.e.}
		\begin{equation}
			g\left(P_{\text{est}}-\sqrt{-\ln(1-p)}N^{-\frac{1}{2}} \: , \: \epsilon\right)<\frac{1}{1+2\epsilon}-\frac{1}{2} \nonumber
		\end{equation}
		where $p$ is the probability of generating successfully the final bit with bias $\epsilon'$ (see Thm. 2 in main text).
	
	\item The QB and the SoR fulfill the Markov assumption, so that, for any value of $e$
		\begin{equation}
			\label{eq:markovass}
			P(A_1,...,A_N,h_1,...,h_n|\vec{X},e)=P(A_1,...,A_N|\vec{X},e)\times P(h_1,...,h_n|\vec{X},e)
		\end{equation}

\end{enumerate}

The first step in the proof is to bound how much the observed average success probability $P_{\text{est}}$ deviates from the real average probability of success $P_{\text{ave}}:=\frac{1}{N}\sum_{j=1}P_{s}^{j}$, where $P_{s}^{j}$ is the success probability of the $j$-th use of the tripartite box in the QB. Note that $P_{s}^{j}$ may indeed depend on the previous bits generated in the QB for the $j-1$ previous runs of the QB. That is, $P_s^j$ should be understood as the probability of success of the $j$-th round conditioned on  $A_1,...,A_{j-1},\vec{X}_1,...,\vec{X}_{j-1},h_1,...,h_N,e$.

We can follow the reasoning from \cite{rndexp2,rndexp5,rndexp6} and use Azuma-Hoeffding inequality to establish that
\be
	\label{Azuma}
	\text{Prob}(P_{\text{ave}}\leq P_{\text{est}}-x)\leq \exp\left(-\frac{1}{2}x^2 N \right).
\ee

Let us define $P^*$ as the success probability fulfilling $g(P^*,\epsilon)=\frac{1}{1+2\epsilon}-\frac{1}{2}$. Then, by \eqref{eq:estsuccess} we have that
\begin{equation}
	P_{\text{est}}>P^*+\sqrt{-\ln(1-p)}N^{-\frac{1}{2}}. \nonumber
\end{equation}
Note that by taking $x=\sqrt{-\ln(1-p)}N^{-\frac{1}{2}}$, one obtains that
\begin{equation}
	\text{Prob}\left(P_{\text{ave}}\leq P^*\right)\leq 1-p, \nonumber
\end{equation}
or equivalently, with probability $p$ it is fulfilled that  $g(P_{\text{ave}},\epsilon)\leq g(P^*,\epsilon)= \frac{1}{1+2\epsilon}-\frac{1}{2}$. For ease of presentation in the following we assume that this latter condition is fulfilled. Results obtained under this assumption can be only be certified to occur with probability $p$. Let us denote the bias of each of the runs of the QB by
\begin{equation}
	\label{eps_j}
	\frac{1}{2}-\epsilon_j \leq P(A_j|\vec{a_j},\vec{X}_{j-1},e)\leq \frac{1}{2}+\epsilon_j
\end{equation}
We can then use the concavity of the function $g$ to derive
\be
	\label{w1-d}
	\epsilon_{\text{ave}}=\frac{1}{N}\sum_{j=1}^N \epsilon_j \leq \frac{1}{N}\sum_{j=1}^N g(P_s^j,\epsilon) \leq g(P_{ave},\epsilon) < \frac{1}{1+2\epsilon}-\frac{1}{2}.
\ee

Let us note that this bound on the average bias allows one to upper bound the probability of obtaining a certain combination of $A_1,...,A_n$. That is,
\be
	\label{eq:boundA}
	P(A_1,...,A_n|\vec{X},e)\leq \prod_{j=1}^N \left(\frac{1}{2}+\epsilon_j\right)\leq \left(\frac{1}{2}+\epsilon_{\text{ave}}\right)^N.
\ee
Equivalently, by assumption SoR produces bits $h_1,\ldots,h_n$ that fulfill
\be
	\label{eq:boundh}
	P(h_1,...,h_n|\vec{X},e)\leq \left( \frac{1}{2} +\epsilon\right)^N.
\ee
This together with \eqref{eq:markovass} will be sufficient to distill a final random bit of an arbitrarily small bias. Note that \eqref{eq:markovass} implies that conditioned on any value of $\vec{X},e$, the sources producing $A$'s and $h$'s are independent. Results on distillation of random bits from independent imperfect random sources were first derived in Ref. \cite{SV}. Here we employ a more recent distillation method that is best suited for our scenario.

\begin{lem}
	\cite{chor} Let us define a $(N,b)$-source one that produces $N$ bits $s_1,...,s_N$ such that $P(s_1,...,s_N|k)\leq2^{-b}$. Consider two independent sources $(N,b_1)$ and $(N,b_2)$ producing bits $s_1,...,s_N$ and $p_1,...,p_N$ respectively. The independence condition reads
	\be
		P(s_1,...,s_N,p_1,...,p_N|k)=P(s_1,...,s_N|k)\times P(p_1,...,p_N|k). \nonumber
	\ee
	Then, the inner product $y=\bigoplus_{i=1}^N s_i \cdot p_i$ is an $\epsilon'$-free bit --that is $\frac{1}{2}-\epsilon' \leq P(y|k)\leq\frac{1}{2}+ \epsilon'$-- if
	\begin{equation}
		b_1+b_2\geq N+2+2\log_2\frac{1}{ \epsilon'}. \nonumber
	\end{equation}
\end{lem}

Previous lemma can be straightforwardly applied to our scenario. As derived above, the QB and SoR are two independent sources when conditioned on $\vec{X},e$ with bounds \eqref{eq:boundA} and \eqref{eq:boundh}. This implies that they are $(N,b_1)$ and $(N,b_2)$ sources with
\begin{eqnarray}
	b_1 &=& N\log_2\left(\frac{1}{2}+\epsilon_{\text{ave}}\right)^{-1}\\
	b_2 &=& N\log_2\left(\frac{1}{2}+\epsilon\right)^{-1}
	\nonumber
\end{eqnarray}
Hence, the bit $y=\bigoplus_{i=1}^N A_i \cdot h_i$ is a $\epsilon'$-free bit ( \emph{i.e.} it fulfills $\frac{1}{2}-\epsilon'\leq P(y|\vec{X},e)\leq \frac{1}{2}+\epsilon'$) as long as
\be
	N\log_2\left( \left(\frac{1}{2}+\epsilon_{\text{ave}}\right)^{-1}\left(\frac{1}{2}+\epsilon\right)^{-1}\right) \geq N+2+\log_2 \frac{1}{\epsilon'} \nonumber
\ee
This can be achieved for any value of $\epsilon'$ by increasing the value of $N$, if $\log_2\left( \left(\frac{1}{2}+\epsilon_{\text{ave}}\right)^{-1}\left(\frac{1}{2}+\epsilon\right)^{-1}\right)> 1$ or equivalently if $\epsilon_{\text{ave}}<\frac{1}{1+2\epsilon}-\frac{1}{2}$, which is satisfied with probability $p$, as derived above.

\section{Appendix B: Bounds on amplification with trivial hashing function}

As described in previous sections, our protocol for randomness amplification makes use of bits $h_1,\ldots,h_N$ taken from SoR to construct a hashing function. This method distinguishes our protocol from the one in \cite{Amp1} where a trivial hashing function is applied to obtain the final random bit. That is, the final random is assigned to one of the outputs $A_i$ where $i$ is chosen by use of the SoR. In \cite{Amp2} a deterministic hashing function is applied, nonetheless their protocol allows for amplification up to $\epsilon'=\frac{1}{4}$ from arbitrarily deterministic sources by applying  a trivial hashing function. In this section we show that such trivial hashing function is useful for amplification only in the noise free case.
If the quantum resources available are not perfect, every set of measurements on each entangled state has a probability larger than zero of not fulfilling the conditions of the nonlocal game. Hence, for large values of $N$, the estimated success probability will converge to $P_{est}=1-\kappa$, with $\kappa > 0$. Next, we show that for every value of $\kappa > 0$, if a trivial hashing function is applied to the outputs $A_1,\ldots,A_N$, no randomness amplification is possible for $\epsilon>0.345$. This is shown by constructing an explicit attack that: (i) provides a value of $\kappa\rightarrow 0$ in the limit of large N, and (ii) the eavesdropper possess a classical variable $\lambda$ perfectly correlated with the final random bit $y$. The attack is defined independently of the nonlocal game or quantum states employed.

The attack is defined as follows:

\begin{itemize}
\item The eavesdropper has to provide $N\equiv 2^ k$ nonlocal boxes. The classical variables used as inputs of the nonlocal boxes are chosen by tossing a fair coin without any intervention by the eavesdropper.
\item When the trivial hashing function is applied, $k$ bits $h_1,...,h_k$ are provided by the SoR, whose bias is controlled by the eavesdropper. These bits are used to choose one of the $N$ nonlocal boxes whose output $A_{h_1,...,h_k}$ would be chosen to be the final bit $y$ of the protocol. That is, each nonlocal box is labeled by a value of the string $h_1,...,h_k$.
\item The eavesdropper chooses a string $\tilde{h}_1,\ldots,\tilde{h}_k$. The bias on the SoR prepared by the eavesdropper is such that
\begin{equation}\label{eq:probhs}
P(h_j=\tilde{h}_j|h_1,\ldots,h_{j-1},h_{j+1},\ldots,h_k)=P(h_j=\tilde{h}_j)=\frac{1}{2}+\epsilon
\end{equation}
for every value of $h_1,\ldots,h_{j-1},h_{j+1},\ldots,h_k$ and $j$. That is, each bit is independently and identically biased.
\item Let us denote by $G_t$ the group of nonlocal boxes labeled by a string $h_1,...,h_k$ that fulfills $\sum_{j=1}^{k} h_j \oplus \tilde{h}_j=t$. As such, $t\in \{0,\ldots,k\}$ and $G_t$ contains $\binom{k}{t}$ elements. Boxes in the groups $G_t$ with $t \in \{0,\ldots, \lfloor (\frac{1}{2}-\alpha)k \rfloor \}$, where $0<\alpha<\frac{1}{2}$, are chosen by the eavesdropper to be classical deterministic boxes. In order to simplify further calculations let us take the worst case scenario where the classical deterministic have null success probability in the nonlocal game. The rest, are perfect quantum states with unit success probability in the nonlocal game.
\end{itemize}

Given such strategy,
\begin{eqnarray}
\kappa &\leq& \frac{\sum_{t=0}^{ \lfloor (\frac{1}{2}-\alpha)k \rfloor}  \binom{k}{t} }{2^k}
\end{eqnarray}
which tends to zero with $k$ tending to infinity. This ensures that  condition (i) above is fulfilled.

When the variables $h_1,\ldots,h_k$ are produced by the SoR, if a box from $G_t$ with $t \in \{0,\ldots, \lfloor (\frac{1}{2}-\alpha)k \rfloor \}$ is chosen, then, since these boxes are classical and deterministic, the eavesdropper possess a classical variable perfectly correlated with the final bit. Let us now denote that probability of choosing one of such classical nonlocal boxes by $P_{\text{attack}}$. Note that, according to \eqref{eq:probhs}, each box in a group $G_t$ is chosen with probability $ \left(\frac{1}{2}+\epsilon \right)^{k-t} \left(\frac{1}{2}-\epsilon\right)^t$. Hence,
\begin{eqnarray}
1-P_{\text{attack}} &=&  \sum^{k}_{t= \lfloor (\frac{1}{2}-\alpha)k \rfloor +1 } \left(\frac{1}{2}+\epsilon \right)^{k-t} \left(\frac{1}{2}-\epsilon \right)^t  \binom{k}{t}  \leq \sum^{k}_{t=  \lfloor (\frac{1}{2}-\alpha)k \rfloor +1 } \left(\frac{1}{2}+\epsilon \right)^{k-t} \left(\frac{1}{2}-\epsilon\right)^t \left(\frac{ke}{t} \right)^t\\
&\leq & \sum^{k}_{t=  \lfloor (\frac{1}{2}-\alpha)k \rfloor +1 } \left(\frac{1}{2}+\epsilon \right)^{k-t} \left(\frac{1}{2}-\epsilon\right)^t \left(\frac{ke}{(\frac{1}{2}-\alpha)k} \right)^t = \left(\frac{1}{2}+\epsilon \right) ^k \sum^{k}_{t=  \lfloor (\frac{1}{2}-\alpha)k \rfloor +1 }\left( \frac{  (\frac{1}{2}-\epsilon)e }{(\frac{1}{2}+\epsilon)(\frac{1}{2}-\alpha)}\right) ^ t.
\end{eqnarray}
which tends to zero for $N,k\rightarrow \infty$ if $\left( \frac{  (\frac{1}{2}-\epsilon)e }{(\frac{1}{2}+\epsilon)(\frac{1}{2}-\alpha)}\right) \leq 1$. This is fulfilled for $\alpha \rightarrow 0$ if $\epsilon \geq \frac{2e-1}{4e+2} \approx 0.345$.

If, instead of a single round, SoR is used to choose a constant number $K$ of them, the same condition on $\epsilon$ holds as the probability of choosing at least one nondeterministic round is $\left(1 -P_{\text{attack}}\right)^K$ which also tends to zero for $N,k\rightarrow \infty$ when $\epsilon \geq \frac{2e-1}{4e+2} \approx 0.345$.

\end{widetext}
\end{document}